# Knowledge management for enterprises

(Wissensmanagement für Unternehmen)

Wolfgang Eiden, January 2000

feedback@wolfgang-eiden.de

http://www.wolfgang-eiden.de

**Abstract:** Although knowledge is one of the most valuable resource of enterprises and an important production and competition factor, this intellectual potential is often used (or maintained) only inadequate by the enterprises. Therefore, in a globalised and growing market the optimal usage of existing knowledge represents a key factor for enterprises of the future. Here, knowledge management systems should engage facilitating. Because geographically far distributed establishments cause, however, a distributed system, this paper should uncover the spectrum connected with it and present a possible basic approach which is based on ontologies and modern, platform independent technologies. Last but not least this attempt, as well as general questions of the knowledge management, are discussed.

**ACM-Classification:** H.3.0, I.2.4

**Zusammenfassung:** Obwohl Wissen eine der wertvollsten Ressourcen von Unternehmen ist und einen wichtigen Produktions- und Wettbewerbsfaktor darstellt, wird dieses intellektuelle Potential oftmals von den Unternehmen nur unzureichend genutzt und gepflegt. Von daher stellen in einem globalisierten und wachsenden Markt die optimale Nutzung von vorhandenem Wissen Schlüsselfaktoren für die Unternehmen der Zukunft dar. Hierbei sollten Wissensmanagementsysteme unterstützend eingreifen. Da geographisch weit verteilte Niederlassungen allerdings ein verteiltes System bedingen, soll dieses Paper das damit verbundene weite Spektrum aufgedecken und einen möglichen Lösungsansatz, der auf Ontologien und modernen, plattformunabhängigen Technologien basiert, vorstellen. Im Anschluß werden sowohl dieser Ansatz, als auch generelle Fragestellungen des Wissensmangements einem allgemeinen Diskurs unterzogen.

# 1 Einleitung

Bedingt durch den wachsenden und sich ständig verändernden Weltmarkt stehen die Unternehmen in der heutigen Zeit zur Erhaltung der Konkurenzfähigkeit mehr denn je vor der Herausforderung des effektiven Managements von Wissen. Bei einer schnelleren und präziseren Verfügbarkeit von Wissen kann das Unternehmen bei Entwicklungs- und Entscheidungsprozessen in einem wesentlich kürzeren Zeitraum reagieren. Insbesondere bei hoch innovativen und qualifizierten Stellen ist die Arbeitsleistung jedes einzelnen Mitarbeiters oder Teams und somit der ganzen Organisation stark vom vorhandenen Wissen beziehungsweise von der schnellen und intuitiven Abfrage von verfügbaren Wissen abhängig.

Daher stellt sich für die Unternehmen unmittelbar die Frage, wie Wissen trotz innerer und äußerer Einflüsse bewahrt, gewonnen und zur Verfügung gestellt werden kann, um eine maximale Effektivität bei möglichst hoher Flexibilität zu erreichen:

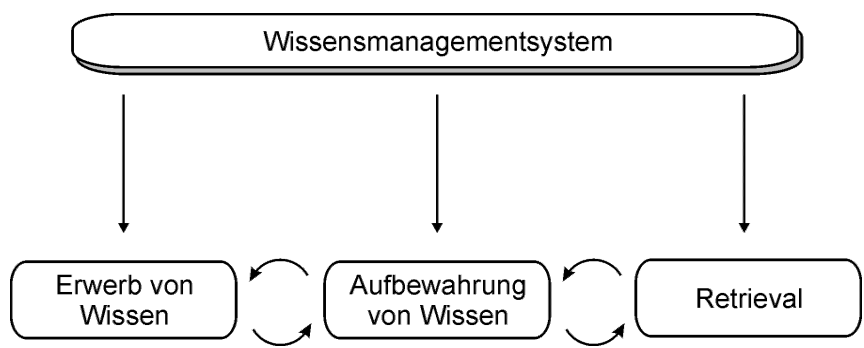

Einflüsse hierbei sind beispielsweise Personalveränderungen (Zuwachs, aber auch Verlust von Mitarbeitern) und die sich schnell ändernde Marktlage. Über die Akquisition und Aufbewahrung von Wissen hinaus sollten die Wissens-managementsysteme sowohl einen wirkungsvollen Informationsaustausch, als auch ein effzientes Retrieval ermöglichen [6]. Hierbei sind die, nur für ein spezielles Teilgebiet entwickelten Systeme zwar höchst effektiv, dafür aber extrem situations-spezifisch. Diese Systeme, die meistens als fertige Produkte direkt zu erwerben sind, besitzen somit nur einen geringen Wert für andere Geschäfts-situationen [2]. Bei Systemen, die eine Vielfalt von Geschäftssituationen abdecken, sollte es sich aber eher um individuell anpaßbare und flexible Systeme handeln, da das effektive Management von Wissen in vielen Bereichen von sozialen Faktoren, Geschäfts-praxen und Informationstechnologien abhängig ist (vgl. [2], [6]). Man denke hierbei beispielsweise an HelpDesk-Applikationen (beispielsweise auf Call-Center-Basis): Damit die Mitarbeiter nicht durch die Eintragung und die Organisation von Wissen gestört werden, sollte das System diese bei seiner Arbeit begleiten, beobachten und automatisch interessante Informationen sammeln und sichern (Hierbei ist die Beobachtung aber nicht im, von George Orwell's visionären Roman 1984 geprägten, "Big brother is watching you" - Sinn gemeint). Daher sind neben der Integration in die Arbeitsumgebung die Anpassungsfähigkeit und die Selbstorganisation von elementarer Wichtigkeit. Wissensmanagementsysteme sollten darüber hinaus selbstständig wissenswerte Informationen anbieten und dabei individuelle Informationsfilterung erlauben [1].

## 2 Ausgangssituation

Im Allgemeinen liegt in den Firmen Wissen oftmals in heterogenen Repräsentationen vor:

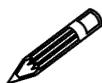 Protokolle

Datenbanken

Versuchsdaten

Dokumentationen

Notizen

Dokumente

Gesetzesbestimmungen

Normen

Unstrukturierte Daten, formal spezifiziertes Wissen (wie beispielsweise Arbeitsabläufe und formale Entscheidungsregeln) und informell gegebenes Wissen (wie beispielsweise Notizen und Dokumentationen) stehen meistens zusammen-hanglos nebeneinander und sind auch nicht immer elektronisch verfügbar [6].

Aufgrund der weitgehend internationalen Ausrichtung der Unternehmen und globalen Infrastrukturen sind die, an einem konkreten Entwicklungsprozeß beteiligten, Personen oder Organisationen oftmals auf mehrere, geographisch weit entfernte Standorte verteilt. Insbesondere hier ist zur Erreichung einer engen und effektiven Kooperation mit Lieferanten, Firmenkunden und Servicediensten die Verwendung von verteilten Wissensmanagementsystemen notwendig [6]. Man denke beispielsweise an einen Projektleiter der ein Entwicklungsteam zusammenstellen möchte welches über mehrere Spezialgebiete hinaus standortübergreifend operieren soll. Neben der Suche nach kompetenten Ansprechpartnern für den Meinungs- und Erfahrungsaustausch bedingt dies selbstverständlich auch die Suche für geeignete Teammitglieder. Oftmals können die verantwortlichen Personen aufgrund der Betriebsgrößen- und organisation nicht persönlich alle geeigneten Personen kennen oder über die einzelnen Kompetenzen und Fähigkeiten Bescheid wissen. Darüber hinaus wissen die Mitarbeiter eines Unternehmens oftmals nicht, ob (und gegebenenfalls wo) überhaupt nützliche Informationen zur Verfügung stehen und die Suche nach Informationen kann somit eine Menge Zeit und damit auch viel Geld kosten.

## 3 Ziele und Leitlinien

Motiviert durch die geschilderte Ausgangssituation sollte also ein Wissensmanagementsystem eine kontinuierliche, räumlich verteilte und nicht ausschließlich auf eine Aufgabe fixierte Verwaltung von unternehmensspezifischen Wissensbeständen, Informationen und Daten ermöglichen. Es sollte darüber hinaus auch die Wartung (beispielsweise Korrektur, Aktualisierung, Ergänzung und Löschung von Informationen) und optimale Ausnutzung von Wissen ermöglichen. Eine wichtige Rolle spielt dabei die Aufdeckung von Wissen (Knowledge refinement). Hierbei ist die Entwicklung von neuem Wissen im Allgemeinen ein iterativer Prozeß, der durch Auswahl von vorhandenem Wissen und deren Variation beziehungsweise Kombination geprägt wird:

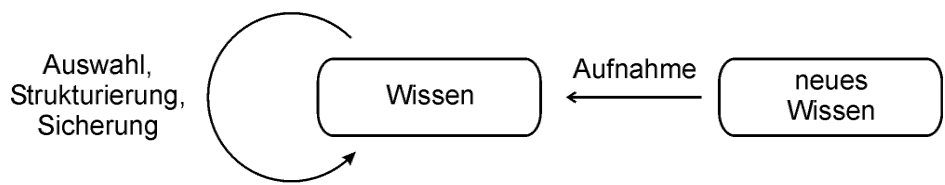

Dies kann beispielsweise durch die Explizitierung von versteckten Zusammenhängen geschehen: Ist beispielsweise jemand Mitarbeiter eines Projektes das von einer bestimmten Organisation in Auftrag gegeben wurde, so kann daraus geschlossen werden, dass diese Person mit der Organisation auch Erfahrungen gesammelt hat:

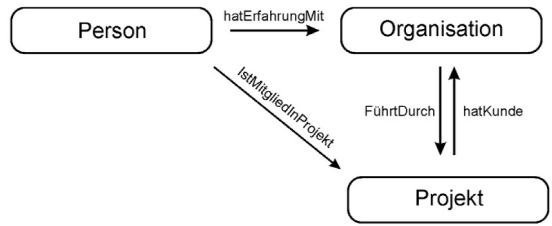

Die Gewinnung von explizitem Wissen kann beispielsweise bei einer geschickten Wissensrepräsentation mit Ontologien durch geeignete Inferenzmechanismen erreicht werden.

Bei der Verflechtung von Wissen sind auch dessen Zusammenhänge und Wechselbeziehungen von Bedeutung. So können beispielsweise aufgrund eines Sitzungsbeschlusses bei dem konkrete Analysen ausschlaggebend waren, wichtige Entscheidungen zur weiteren Entwicklung getroffen werden. Nur durch Erkennung von Zusammenhängen ist demzufolge ein intelligentes Retrievalverfahren realisierbar bei dem auch Fragen wie "Gibt es an unserem Institut eine Person die programmieren kann, sich irgendwann einmal mit künstlicher Intelligenz und Ontologien beschäftigt und Erfahrungen mit einer konkret vorgegeben Organisation gesammelt hat ?" beantwortet werden können. Daher erlangt zunächst der Aufbau, die Sammlung und die Sicherung des unternehmensspezifischen Kernwissens einen enormen Stellenwert. Diesen Prozeß bezeichnet man auch als Knowledge Gathering. Dazu gehört auch bereits erfaßtes Wissen, welches aufgrund mangelnder Referenzen oder fehlender beziehungsweise ungenügender Retrievalmöglichkeiten (qualitatives Retrieval versus quantitatives Retrieval) nur unzureichend genutzt wird [1]. Ein wichtiger Aspekt ist also in diesem Zusammenhang die bessere Erschließung von stabilem und wertvollem, bereits vorhandenem Wissen.

Die Bereitstellung, die Organisation und die Strukturierung des erworbenen Wissens (Knowledge organization and structuring) ist dann selbstverständlich ein weiterer wichtiger Punkt. Die Wissensrepräsentation sollte hierbei sowohl eine flexible

Strukturierung von Wissen, als auch rasche Anpassungsmöglichkeiten an neue und sich ändernde Anforderungen erlauben [6]. Ist ein gewisser Wissensvorrat einmal vorhanden, so sollte dieser zur Optimierung des Retrievals und des Wissenstransfers zunächst veredelt und dann derart verteilt werden (Knowledge distribution), dass es auch wirklich denjenigen zur Verfügung steht, die es benötigen.

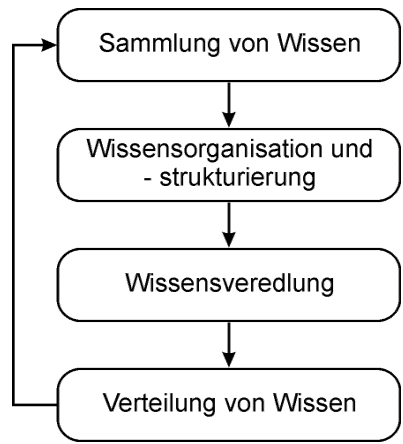

Da man nicht mehr davon ausgehen kann, dass firmenweit nur auf einer Computerplattform gearbeitet wird, sollte das System dabei plattform-unabhängig und standortübergreifend verfügbar sein. Daher scheint aus heutiger Sicht (auch unter dem Aspekt der Verwendung zukunftsorientierter Technologien) die Nutzung des Internets beziehunsweise Intranets mit seinen mittlerweile weltweit anerkannten Informationsdiensten (wie beispielsweise World-Wide-Web oder electronic-mail) die logische Konsequenz. Neben der Plattformunabhängigkeit ist hierbei die Nutzungsmöglichkeit des Angebotes ohne die Installation spezieller Software ein weiterer Vorteil.

# 4 Lösungsansatz

Aufgrund der Dezentralisation, der Vielfalt und der unterschiedlichsten Strukturierungen von Wissen ergeben sich oftmals uneinheitliche Wissens-repräsentationen. Deshalb sollte vor allem die Konzeptionalisierung von Wissen auf Basis einer gemeinsamen und firmenweit akzeptierten Ontologie (als Hintergrund-theorie von Wissen) ermöglicht werden. Das Bestreben hierbei ist unter anderm die Entdeckung von impliziten Wissen, also von Wissen, welches nicht explizit niedergeschrieben ist, aber aus dem generellen Wissen der Ontologie hergeleitet werden kann. Die informationstechnischen Aspekte des Wissensmanagement bestehen also im Wesentlichen sowohl aus einer geeigneten Strukturierung von Wissen auf Basis einer Ontologie, als auch aus entsprechende Abfrage-möglichkeiten und Herleitungsmechanismen von Wissen. Daher beschränkt sich der hier vorgestellte Lösungsansatz auf die Betrachtung dieser, mit dem internet-basierten Wissensmanagement verbundenen, informationstechnologischen Aspekte. Die generellen Ideen dieses Ansatzes werden anhand eines konkreten Beispiels, nämlich der Organisationsstruktur einer Forschungsgruppe, verdeutlicht:

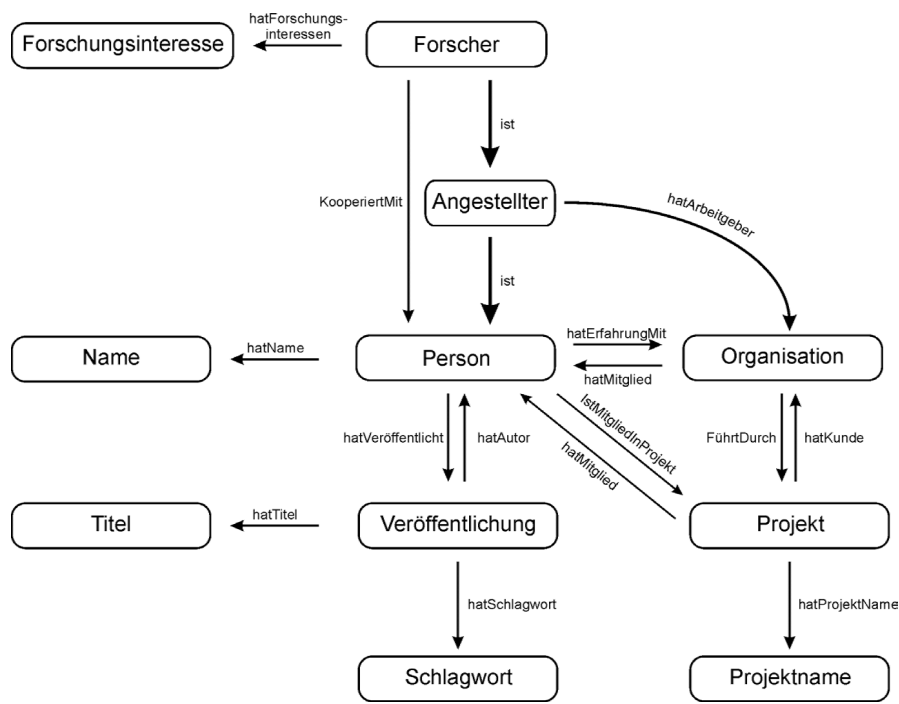

Aufgrund der Einfachheit des gewählten Beispiele mögen manche der nun folgenden Ausführungen und Zusammenhänge zunächst trivial erscheinen. Diese Zusammen-hänge und Wechselbeziehungen müssen aber ersteinmal einem Computersystem (auf anständige Art und Weise) vermittelt werden.

Die Forscher einer Gruppe sind neben der Tatsache, dass sie Angestellte der Forschungsgruppe sind, zunächst einmal sicherlich auch Personen. Sie haben Forschungsinteressen, können in Projekten mitarbeiten und Veröffentlichungen schreiben. Bereits in diesem kleinen Modell kann sich bei einer konkreten

Instanziierung implizites Wissen verbergen: Ist beispielsweise eine Person Mitglied in einem Projekt, dann kann man im Allgemeinen davon ausgehen, dass diese auch mit der auftraggebenden Organisation Erfahrungen gesammelt hat.

## 4.1 Spezifikation von Ontologien

Die gemeinsame Nutzung von Ontologien setzt eine adäquate Sprache zur Formalisierung beziehungsweise Spezifikation voraus.

### 4.1.1 Framelogik

Prädestiniert hierfür ist beispielsweise die Framelogik [7], da sowohl attributierte Klassen, als auch Logikregeln definiert werden können.

**Klassen und Attribute**

Eine Klasse beschreibt einen Objekttyp. Sie definiert die Menge der unterschiedlichen Objekte, die sich aus einem Satz identischer Eigenschaften bilden lassen. Hierbei wollen wir unter Objekten Entitäten verstehen, die von anderen Entitäten eindeutig unterscheidbar sind, eine eindeutige Identität besitzen und mit den zu modellierenden Gegenständen korrespondieren. Durch das Prinzip der Vererbung können neue Klassen durch Wiederverwendung bereits existierender Klassen gebildet werden. Die Syntax hierfür lautet

```
Unterklasse :: Oberklasse.
```

Vordefiniert ist die Klasse `Object`, die als Vaterklasse fungiert. Somit kann beispielsweise in Anlehnung an das Modell folgende Hierarchie definiert werden:

```
TObject.
        TOrganisation :: TObject.
        TPerson :: TObject.
                TAngestellter :: TPerson.
                        TForscher :: TAngestellter.
        TProjekt :: TObject.
        TVeroeffentlichung :: TObject.
```

Die Charakterisierung einer Klasse wird durch zusätzliche klassenspezifische Definition von Attributen unterschiedlichsten Typs erreicht, wobei bei der Vererbung die Unterklasse alle Attribute der referenzierten Oberklasse übernimmt. Neben einfachen Klassen wie `STRING` können zur Definition von Attributen auch die selbstdefinierten Klassen verwendet werden (diese dürfen aber verständlicherweise nicht vom Typ einer Unterklasse der zu definierenden Klasse sein). Hierbei wird die Syntax

```
Klasse[Attribut ==> Typ]
```

verwendet. Wird die Klasse Klasse nun instanziiert, so ist der Wert des Attributes Attribut der Klasse eine Instanz der Klasse Typ. Hierbei ist zu beachten, dass Attribute immer als Sets zu betrachten sind, bei der Instanziierung also auch mehrfach belegt werden können (dies macht natürlich nicht notwendigerweise für alle Fälle Sinn). In Anlehnung an das Modell könnte dann beispielsweise folgendes definiert werden:

```
TOrganisation[
    HatName ==> STRING
    HatMitglied ==> TPerson
    FuehrtDurch ==> TProjekt
...].

TPerson[
    HatName ==> STRING
    HatVeroeffentlicht ==> TVeroeffentlichung
    IstMitgliedInProjekt ==> TProjekt
    HatErfahrungMit ==> TOrganisation
...].

TAngestellter[
    HatArbeitgeber ==> TOrganisation
...].

TForscher[
    HatForschungsinteressen ==> STRING
    KooperiertMit ==> TPerson
...].

TProjekt[
    HatProjektname ==> STRING
    HatMitglied ==> TPerson
    HatKunde ==> TOrganisation
...].

TVeroeffentlichung[
    HatAutor ==> TPerson
    HatTitel ==> STRING
    HatSchlagwort ==> STRING
...].
```

Durch die so definierten Klassen wird das Modell vollständig wiedergegeben. Wie bereits aus diesen Definition ersichtlich wird, besteht ein wesentlicher Vorteil der Vererbung und der Attributierung in der klareren Strukturierung des betrachteten Wissensgebietes.

**Regeln**

Eine wesentliche Eigenschaft von Framelogik ist dieMöglichkeit der Definition von Folgerungsund Äquivalenzregeln. Gegenüber dem Modell in dem nur durch die menschliche Intuition und den Erfahrungshorizont implizites Wissen gegeben ist, kann dieses hier formal niedergeschreiben werden. Beispielsweise wird durch die Folgerungsregel

```
FORALL PE1, PR1, OR1
    PR1 : TProjekt
    [HatMitglied ->> PE1 ; HatKunde ->> OR1]
->
    PE1 : Person
    [HatErfahrungMit ->> OR1].
```

der bereits kennengelernte Zusammenhang ausgedrückt. Neben Folgerungsregeln werden auch Regeln zur Definition von Äquivalenzen zugelassen. Ist beispielsweise der konkrete Autor eines Buches bekannt, so liegt es auch nahe zu behaupten, dass diese Person ebenfalls dieses Buch geschrieben hat. Formal kann dies dann so notiert werden:

```
FORALL PE1, VE1
    VE1 : TVeroeffentlichung
    [HatAutor ->> PE1]
<->
    PE1 : TPerson
    [HatVeroeffentlicht ->> VE1].
```

Oder ein Beispiel für eine Konsistenzbedingung: Wenn ein Forscher mit einem anderen kooperiert, dann gilt dies im Allgemeinen auch umgekehrt (Symmetrie des "Kooperiert- Mit"-Attributes):

```
FORALL PE1, PE2
    PE1 : TForscher
    [KooperiertMit ->> PE2]
<->
    PE2 : TForscher [KooperiertMit ->> PE1].
```

| Klassen | Attribute | Regeln |
|---|---|---|
| TObject[]. <br>   TOrganisation :: TObject. <br>   TPerson :: TObject. <br>     TAngestellter :: TPerson. <br>       TForscher :: TAngestellter. <br>   TProjekt :: TObject. <br>   TVeröffentlichung :: TObject. | TOrganisation[ <br>   HatName ==> STRING <br>   HatMitglied ==> TPerson <br>   FührtDurch ==> TProjekt <br> ...]. <br><br> TPerson[ <br>   HatName ==> STRING <br>   HatVeröffentlicht ==> TVeröffentlichung <br>   IstMitgliedInProjekt ==> TProjekt <br>   HatErfahrungMit ==> TOrganisation <br> ...]. <br><br> TAngestellter[ <br>   HatArbeitgeber ==> TOrganisation <br> ...]. <br><br> TForscher[ <br>   HatForschungsinteressen ==> STRING <br>   KooperiertMit ==> TPerson <br> ...]. <br><br> TProjekt[ <br>   HatProjektname ==> STRING <br>   HatMitglied ==> TPerson <br>   HatKunde ==> TOrganisation <br> ...]. <br><br> TVeröffentlichung[ <br>   HatAutor ==> TPerson <br>   HatTitel ==> STRING <br>   HatSchlagwort ==> STRING <br> ...]. | FORALL PE1, PE2 <br>   PE1 : TForscher <br>   [KooperiertMit ->> PE2] <br> <-> <br>   PE2 : TForscher <br>   [KooperiertMit ->> PE1]. <br><br> FORALL PE1, VE1 <br>   VE1 : TVeröffentlichung <br>   [HatAutor ->> PE1] <br> <-> <br>   PE1 : TPerson <br>   [HatVeröffentlicht ->> VE1]. <br><br> FORALL PE1, PR1, OR1 <br>   PR1 : TProjekt <br>   [HatMitglied ->> PE1, HatKunde ->> OR1] <br> -> <br>   PE1 : Person <br>   [HatErfahrungMit ->> OR1]. |

## Retrieval

Möchte man nun beispielsweise wissen, mit welchen Forschern Herr Mustermann kooperiert, so könnte folgende Anfrage an die Ontologie gestellt werden:

```
FORALL NAME
<-
    PE1 : TForscher
    [HatName ->> "Mustermann" ; KooperiertMit ->> PE2]
and
    PE2 : TForscher [HatName ->> NAME].
```

Das Ergebnis dieser Abfrage liefert die gewünschten Namen, selbst wenn sie von Herr Mustermann nicht explizit angegeben wurden. Notwendig ist dann nur, dass die jeweiligen (nicht angegebenen) Forscher ihre Kooperation mit Herrn Mustermann festgehalten haben.

Diese Art der Auswertung von Suchanfragen geht weit über das Konzept stichwort-basierter Suchmaschinen hinaus. Beim Retrieval des ontologiebasierten Konzept handelt es sich um ein auf strukturierten Daten beziehungsweise auf geeignet annotierten Daten betriebenes Retrieval. Pilotprojekte wie der Ontobroker der Universität Karlsruhe [2] zeigen hierbei erste Erfolge.

## 4.1.2 Dokumentenannotation

Basierend auf einer aufgebauten Ontologie muß das konkrete Wissen (als Instanziierung der Ontologie) dem System natürlich noch zur Verfügung gestellt werden. Dies kann bei internetbasierten Wissenmanagementsystemen beispielsweise durch eine zentrale Registrationsstelle geschehen, an die die Informationen (beziehungsweise deren URLs) weitergemeldet werden. Leider eignet sich die Hypertext-Markup-Language HTML nur in unbefriedigendem Maße als Beschreibungssprache einer konkreten Ontologieinstanziierung, da zur Strukturierung nur die, in der HTML-Document Type Definition definierten, Tags verwendet werden dürfen. Durch Verwendung der Extensible Markup Language (XML) können diese Probleme allerdings umgangen werden, da damit die Strukturen und Inhalte von Dokumenten präzise beschrieben werden können. Die Verwendung von XML als Sprache wird vom World Wide Web Consortium aufgrund der zu erwarteten hohen Akzeptanz und der Erweiterungsmöglichkeit (ohne dass Spezifikationsänderungen nötig werden) vorgeschlagen.

## 4.1.2.1. XML

Bei der die Extensible Markup Language handelt es sich um einen neuen Internetstandard, der den Austausch, die Nutzung und die Wiederverwendung von Daten in unterschiedlichen Hard- und Softwareumgebungen ermöglichen soll. Hierbei basiert XML auf einer Trennung von Inhalt, Struktur und Layout von Daten beziehungsweise Dokumenten. Mit der Extensible Markup Language können Strukturen und Inhalt von Dokumenten aber so präzise beschrieben werden, dass es letztlich nicht mehr unbedingt notwendig ist, die zum Verständnis und der Weiterverarbeitung von Daten notwendigen Informationen, den Standard, fest in die Anwendung zu integrieren (vgl. [4]). Vielmehr besteht die Vision darin, den auszutauschenden Daten die zu ihrer Nutzung notwendigen Informationen mitzugeben. Neben der Unterstützung des Dokumentenaustausches ermöglicht XML allgemein die flexible Wiederverwendung von Daten und ist die Grundlage für eine Verwendung von Metadaten, was unter anderem zu einer erhöhten Interoperabilität unterschiedlicher Anwendungen und mächtigen Retrievalmöglichkeiten führen kann.

**Document Type Definitions**

Im Gegensatz zu HTML erlaubt XML beispielsweise durch Document Type
Definitions die Definition eigener Tags und ist von daher erweiterbar. Im Beispiel

```
<!ELEMENT TPerson (TName, TEmail, TTelefon)>
<!ELEMENT TName (TVorname, TNachname)>
<!ELEMENT TVorname (#PCDATA)>
<!ELEMENT TNachname (#PCDATA)>
<!ELEMENT TEmail (#PCDATA)>
<!ELEMENT TTelefon (#PCDATA)>
```

wird das Element `TPerson` definiert, das seinerseits zwingend aus dem jeweils
genau einmaligem Auftreten der Elemente `TName`, `TEmail` und `TTelefon`
besteht. Diese Elemente sind in den folgenden Zeilen derart definiert, dass `TEmail`
und `TTelefon` jeweils Text enthalten und `TName` ein Element mit den Elementen
`TVorname` und `TNachname` darstellt. Eine konkrete Instanz dieser Elemente könnte
dann beispielsweise folgendermaßen aussehen:

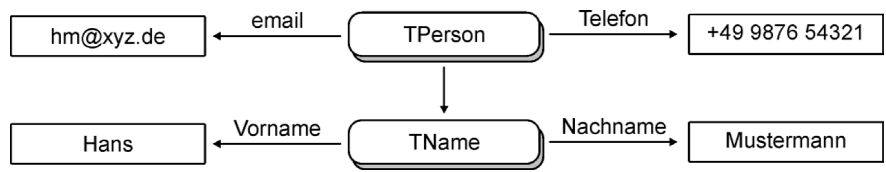

```
<TPerson>
    <TName>
        <TVorname>Hans</TVorname>
        <TNachname>Mustermann </TNachname>
    </TName>
    <TEMail>hm@xyz.de</TEMail>
    <TTelefon>+49 9876 54321</TTelefon>
</TPerson>
```

**Bestandteile von XML**

Aus Benutzersicht sieht ein XML-Dokument zunächst einer HTML-Seite sehr
ähnlich. Entities wie `ä` für ä sind hier wie dort eine wichtige Möglichkeit, zur
Bereitstellung von Zeichen, die nicht notwendigerweise im Standard-ASCII-
Zeichensatz vorhanden sein müssen. Die Extensible Markup Language unterstützt

allerdings von vorneherein als Charactersets sowohl ASCII und ISO-Latin 1 (ISO-8859-1) als auch Unicode (ISO 10.646 UTF 16). Tags kommen sowohl in den bekannten beiden Formen des öffnenden und des schließenden Tags (`<tag>` beziehungsweise `</tag>`), als auch neuerdings als sogenanntes empty tag (`<tag/>`, das gleichzeitig öffnendes und schließendes Tag ist, vor.

Im Gegensatz zu HTML erlaubt XML keine nichtgeschlossenen Tags und ist somit wesentlich restriktiver. Bei der Extensible Markup Language können - wie gesehen - Elemente definiert und sowohl die Ineinander-Schachtelung von Tags, als auch die Form der Daten zwischen öffnenden und schließenden Tags genau festgelegt werden.

Die Definitionen werden formal im sogenannten Prolog eines XML-Dokuments festgelegt (genauer im `<!DOCTYPE>`-Element). Dort kann entweder ein Verweis auf die eigentliche Spezifikation der Dokumentstruktur (was in diesem Zusammenhang sicherlich erwünscht ist) oder aber die vollständige Definition selbst hinterlegt sein. Weitere Details findet man in [4].

### 4.1.2.2. RDF

Obwohl die Extensible Markup Language bereits im Februar 1998 vom W3-Consortium als neuer Standard verabschiedet wurde, bringt die Annotation von Dokumenten mit XML noch einige Probleme mit sich, da in die aktuellen Standardbrowser (Stand: Januar 2000) noch keine vollwertigen XML-Parser integriert sind. Als Übergangslösung schlägt das World Wide Web- Consortium daher das Resource Description Framework [8] vor, das mit der Definition von nur einem neuen Tag (dem `rdf`-Tag) dieses Problem umgeht: Die Definition der Datentypen findet in einer separaten XML-Datei statt und kann dann in einer HTML-Datei innerhalb des rdf-Tags verwendet werden. Sollte die geforderte Integration vollwertiger XML-Parser einmal erfolgt sein, scheint eine Konvertierung von wohlgeformten RDF-Parts nach XML problemlos.

### Properties

Allgemein sind in RDF Ressourcen Objekte, die Informationen enthalten. Die Ressourcen werden durch ihren *Resource Identifier* identifiziert. Sie haben benannte Eigenschaften mit Werten.

Zur eindeutigen Beschreibung der Eigenschaft einer Ressource weist im RDF-Datenmodell ein Pfeil von der Ressource zum Wert der Eigenschaft:

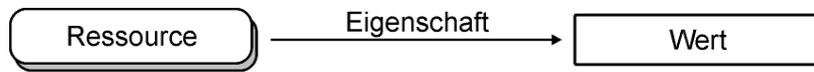

Beispiel:

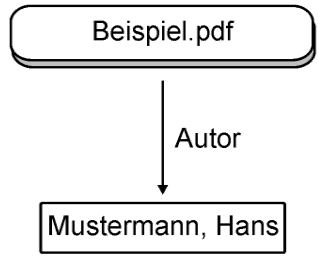

```
<rdf:rdf>
    <rdf:Description about="Beispiel.pdf">
        <s:Autor>
            Mustermann, Hans
        </s:Autor>
    </rdf:Description>
</rdf:rdf>
```

Mehrere Eigenschaften (Properties) einer Ressource werden zu einer Beschreibung (Description) zusammengefaßt. Oftmals sind die Zusammenhänge von komplexerer Ordnung. So ist beispielsweise der Wert des Eigenschaft Autor im Allgemeinen nicht nur eine Zeichenkette, sondern selbst wieder ein attributiertes Objekt, das damit bezüglich dieser Eigenschaft die Rolle einer Ressource übernimmt. Also wird zusätzlich ein Identifier eingeführt, der die Ressource eindeutig beschreibt:

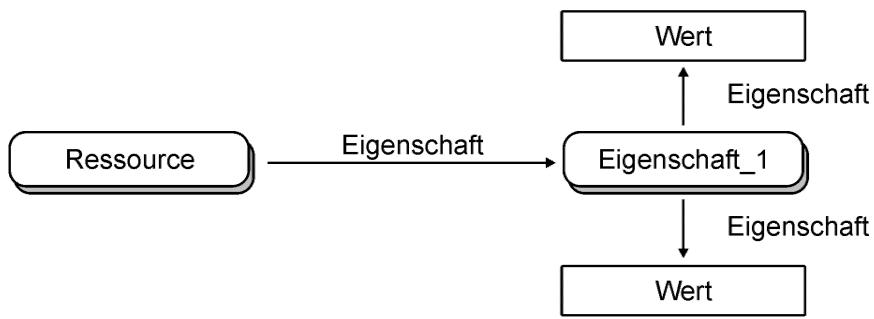

Beispiel:

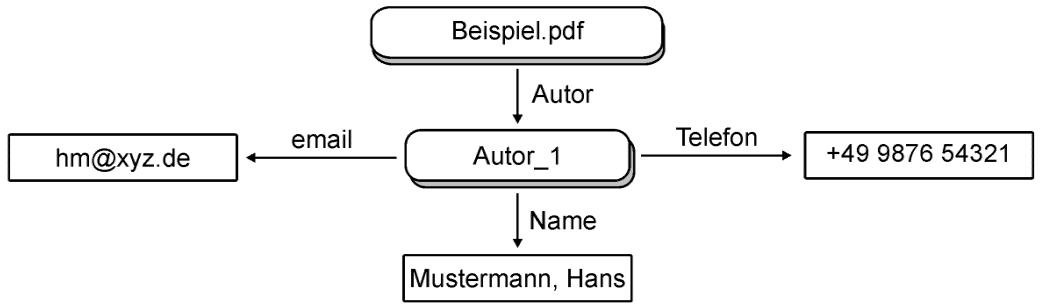

```
<rdf:rdf>
  <rdf:Description about="Beispiel.pdf">
    <s:Autor rdf:ressource="Autor_1"/>
  </rdf:Description>
  <rdf:Description about="Autor_1"/>
    <s:Name>
        Mustermann, Hans
    </s:Name>
    <s:email>
        hm@xyz.de
    </s:email>
    <s:Telefon>
        +49 9876 54321
    </s:Telefon>
  </rdf:Description>
</rdf:rdf>
```

Hier ist Autor_1 ein Resource Identifier, der die Person des Autors eindeutig beschreibt.

**Containers**

Oftmals ist auch die Referenzierung einer Kollektion von Ressourcen notwendig. Dies ist beispielsweise dann der Fall wenn eine Arbeit von mehreren Autoren geschrieben wurde.

Das Resource Description Framework definiert drei Typen von Container-Objekten um solche Zusammenhänge zu modellieren (vgl. [8]):

- *Bag:* Eine ungeordnete Liste von Ressourcen oder Literalen. Bags werden zur Deklaration einer Eigenschaft mit mehreren Werten benutzt, wobei diese in keiner signifikanten Ordnung vorliegen. Dublikate sind erlaubt.

- *Sequence:* Eine geordnete Liste von Ressourcen oder Literalen. Sequences werden zur Deklaration einer Eigenschaft mit mehreren Werten benutzt, wobei die Ordnung von Relevanz ist. Auch hier sind Dublikate erlaubt.

- *Alternative:* Eine Liste von Ressourcen oder Literalen die Alternativen für einen (einzelnen) Wert darstellen.

Um eine Kollektion von Ressourcen zu repräsentieren wird mit der Eigenschaft type eine zusätzliche Ressource gemäß der obigen Liste zur Spezifikation verwendet. Das folgende Beispiel soll dies exemplarisch anhand des Bag-Containers verdeutlichen:

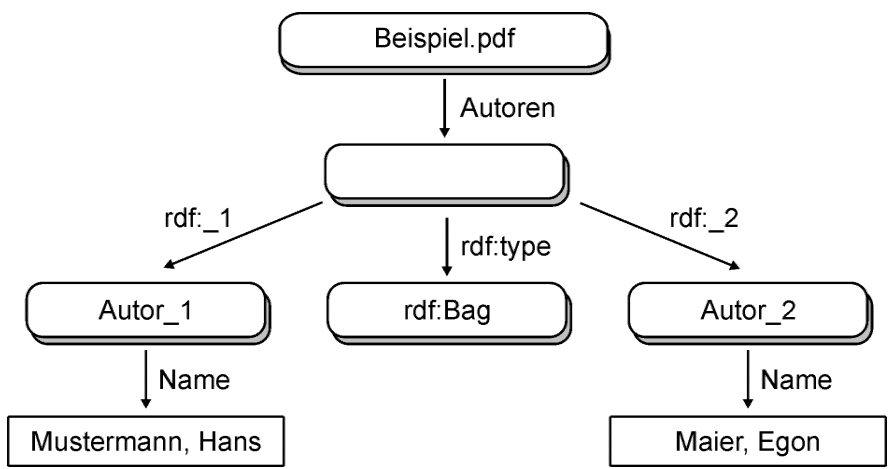

```
<rdf:rdf>
    <rdf:Description about="Beispiel.pdf">
        <s:Autoren>
            <rdf:Bag>
                <rdf:li ressource="Autor_1"/>
                <rdf:li ressource="Autor_2"/>
            </rdf:Bag>
        </s:Autoren>
    </rdf:Description>
    <rdf:Description about="Autor_1"/>
        <s:Name>Mustermann, Hans</s:Name>
    </rdf:Description>
    <rdf:Description about="Autor_2"/>
        <s:Name>Maier, Egon</s:Name>
    </rdf:Description>
</rdf:rdf>
```

**Interpretationsscheatas**

Beim Resource Description Framework wird die Angabe eines Inter-
pretationsschematas (was in unserem Falle mit der Ontologie korrespondiert)
erzwungen: Durch die Qualifizierung des Elementnamens mit einem Name-space-
Präfix [3] wird eine eindeutige Zuordnung des jeweiligen Elements mit dem
korrespondierenden RDF-Schema erreicht. Im Beispiel

```
<rdf:rdf>
    <rdf:Description about="Beispiel.pdf">
        <s:Title>
            Beispielstitel
        </s:Title>
    </rdf:Description>
</rdf:rdf>
```

ist der Namespace-Präfix s eine Referenz auf den verwendeten spezifischen
Namespace-Präfix, der in einer XML Namespace-Deklaration [4] wie beispielsweise

```
xmlns: s="http://description.org/schema/"
```

zu definieren ist.

**Abbreviated Syntax**

Die Einbettung der RDF-Beschreibung in HTML-Code liefert bei Verwendung der
vorgestellten Serialisationssyntax in älteren Browsern leider un-erwünschte
Ausgaben: Die neuen und somit unbekannten Tags werden zwar
(erwünschterweise) nicht dargestellt, jedoch ihre jeweiligen Inhalte. Um dies zu
vermeiden, schlägt das W3-Consortium parallel zu dieser Syntax die Verwendung
der sogenannten Abbreviated Syntax vor. Das Beispiel

```
<rdf:rdf>
    <rdf:Description about="Beispiel.pdf">
        <s:Autoren>
            <rdf:Bag>
                <rdf:li ressource="Autor_1"/>
                <rdf:li ressource="Autor_2"/>
            </rdf:Bag>
        </s:Autoren>
        <rdf:Description about="Autor_1"/>
```

```
        <s:Name>Mustermann, Hans</s:Name>
    </rdf:Description>
    <rdf:Description about="Autor_2"/>
        <s:Name>Maier, Egon</s:Name>
    </rdf:Description>
</rdf:rdf>
```

würde demnach folgende Form annehmen:

```
<rdf:rdf
    xmlns:rdf="http://www.w3.org/1999/02/ 22-rdf-syntax-ns#"
    xmlns:Autoren="http://muster.org/onto#">
    <rdf: Description about="Beispiel.pdf">
        <rdf:Autoren>
            <rdf:Bag>
                <rdf:li ressource="Autor_1"/>
                <rdf:li ressource="Autor_2"/>)
            </rdf:Bag>
        </rdf:Autoren>
    </rdf:Description>
    <rdf:Description about="Autor_1"/>
        <rdf:Name="Mustermann, Hans" />
    </rdf:Description>
    <rdf:Description about="Autor_2"/>
        <rdf:Name="Maier, Egon" />
    </rdf:Description>
</rdf:rdf>
```

Bei Verwendung der Abbreviated Syntax werden auch in älteren Browserversionen (Versionen vor Netscape Communicator 4.5 und Microsoft Internet Explorer 5.0) keine unerwünschten Teile mehr dargestellt.

# 5 Diskussion

Der Erfolg eines Wissensmanagementsystems hängt von vielen Faktoren ab. Einer der Faktoren ist die Akzeptanz durch den Benutzer. Daher muß von den Benutzern zunächst die Wichtigkeit eines Wissensmanagementsystemes erkannt werden und das Bewußtsein wachsen, dass die Verwendung eines solchen Systemes letztendlich eine Zeitersparnis bedeutet.

Das System sollte den Benutzern folglicherweise zumindestens bei den wichtigsten Routinearbeiten unterstützend zur Seite stehen und die Wissensakquisition möglichst wenig Einfluß auf den eigentlichen Arbeitsprozeß haben. Aus diesem Grund sollte sich die Technologie zumindestens größtenteils den Anwendern anpassen (und nicht umgekehrt). Daher sind computerunterstützende Wissensmanagementsysteme vor allem für Betriebe interessant, die ihren betrieblichen Prozeß bereits modelliert und gegebenenfalls computerunterstützt steuern lassen.

Da die geforderte Anpassungsfähigkeit und Selbstorganisation des Systems noch viele offene Probleme aus dem Bereich der Erfassung und Instandhaltung mit sich bringt, ist sowohl ein systematischer Ansatz als auch die Verwendung bereits existierender Technologien zur erfolgreichen Umsetzung dringend notwendig.

Hierzu gehören auch die Entwicklung beziehungsweise Verwendung entsprechender Toolwerkzeuge zur Konstruktion, Verwaltung und Abfrage der Ontologie. So sollte beispielsweise der (leider oft zur Ermüdung und Fehleranfälligkeit tendierende) Annotationsprozess so einfach wie möglich gehalten sein. Dies könnte exemplarisch dadurch geschehen, dass entsprechende Hilfsmittel zur Verfügung gestellt werden, damit Fragmente eines Dokumentes markiert und Informationen durch komfortable und intuitive Termauswahl direkt der Ontologieinstanziierung zur Verfügung gestellt werden. Somit könnten auch Inkonsistenzen aus der Ontologie zwischen dem angezeigten Inhalt und den Anmerkungen weitgehend vermieden werden. Daher ist auch die sorgfältige Planung der Ontologie ein wesentlicher Aspekt, da nachträgliche Änderungen auf der Ontologieebene - gerade bei verteilten Systemen - eventuell tiefgehendene Konsequenzen zur Folge hat.

Ebenfalls als problematisch stellt in diesem Zusammenhang die weltweite Verteilung von Wissen dar: Die (mit den Ontologieänderungen auf Instanziierungsebene verbundenen) notwendigen Aktualisierungsvorgänge sind für einen abgeschlossenen und kompakten Bereich sicherlich durchführbar, jedoch kann die Konsistenz und Zuverlässigkeit bei weltweit verteilten Systemen nie garantiert werden. Man denke hier beispielsweise an das - aus dem Teilgebiet der Nachrichtenkommunikation bekannte - Snapshotproblem: Kernproblem ist hierbei die Festlegung eines konsistenzen Zustandes zu einem beliebig gegebenem Zeitpunkt. Zwar existieren auch hierfür entsprechende Lösungsvorschläge, sind aber ebenfalls wieder mit anderen Problemen und Gefahren (beispielsweise der Gefahr des Deadlocks) verbunden.

Zwar wird es sich wohl erfahrungsgemäß nie vermeiden lassen, dass es verschiedene Versionen einen Ontologie geben wird, dies kann jedoch durch ein geeignetes und integriertes Versions-management kompensiert werden. Ein solches Versionsmangement könnte beispielsweise durch Verwendung von XML-Namespaces [3] unterstützt werden. Das System sollte dabei so beschaffen sein, dass bei Änderungen auf Instanziierungsebene automatisch ein Versions­aktualisierung durchgeführt wird und dass Neueintragungen immer nur in der aktuellsten Version erfolgen dürfen.

Gelingt es die genannten Bedingungen in die Realität umzusetzen, so wird die Verwendung einheitlicher Ontologien und Inferenzalgorithmen sicherlich erhebliche Vorteile gegenüber den herkömmlichen Metadatenbeschreibungungsarten auf­zeigen. So erwartet man beispielsweise, dass das ontologiebasierte Retrieval wesentlich leistungsfähiger als das keywordbasierte sein wird, da dieses nicht wirklich implizites Wissen kennen kann.

Obwohl im Allgemeinen auch die keywordbasierten Suchmaschinen immer "intelligenter" werden (beispielsweise durch Verwendung von Thesauri und Datenbanken), so geschieht dies meist nur innerhalb eines sehr eng begrenzten Rahmens. Dies hängt einerseits damit zusammen, daß - aufgrund der dort fehlenden, übergeordneten Relationen - keine einheitlichen Präsentations­möglichkeiten der Informationen bestehen. Andererseits definiert sich die semantische Ähnlichkeit nicht nur im Sinne von Thesauri ähnlichen Begriffen. Diese kann daher als Vorstufe zum ontologie-basiertem Retrieval angesehen werden.

# 6 Resümee

Auch wenn die menschliche Wissensverarbeitung - die durch Kreativität und Intuition geprägt ist - nicht durch eine formale Wissensverarbeitung zu ersetzen ist, können Wissensmanagementsysteme im Allgemeinen jedoch die Qualität menschlicher Entscheidungen und Problemlösungen durch Anbieten von kontextsensitiven Informationen in erheblichem Maße steigern. Von daher gewinnt das Management von Wissen immer mehr an zentraler Bedeutung und wird langsam aber sicher ein integraler Bestandteil für den Erfolg eines Unternehmens.

# Literatur